\documentclass[lettersize,journal]{IEEEtran}
\usepackage{amsmath,amsfonts,amssymb,amsthm}
\usepackage{algorithmic}
\usepackage{algorithm}
\usepackage{array}
\usepackage[caption=false,font=footnotesize,labelfont=rm,textfont=rm]{subfig}
\usepackage{textcomp}
\usepackage{stfloats}
\usepackage{url}
\usepackage{verbatim}
\usepackage{graphicx}
\usepackage{cite}
\usepackage{xcolor}
\usepackage{multirow}
\usepackage{threeparttable}
\usepackage{tabularx}
\usepackage{array}
\usepackage{amssymb}  
\usepackage{pifont}   

\renewcommand{\arraystretch}{1.3} 
\renewcommand{\multirowsetup}{\centering} 
\begin{document}

\title{Polarforming for Wireless Networks: \\Opportunities and Challenges}

\author{
	Jingze Ding,
	Zijian Zhou,
	Xiaodan Shao,
	Bingli Jiao,
	and Rui Zhang
	\thanks{Jingze Ding is with Peking University, China.}
	\thanks{Zijian Zhou (corresponding author) is with The Chinese University of Hong Kong, China.}
	\thanks{Xiaodan Shao is with University of Waterloo, Canada.}
	\thanks{Bingli Jiao is with Fuyao University of Science and Technology, China, and Peking University, China.}
	\thanks{Rui Zhang (corresponding author) is with The Chinese University of Hong Kong, China, Shenzhen Research Institute of Big Data, China, and National University of Singapore, Singapore.}
}
\maketitle

\begin{abstract}
Polarforming emerges as a promising technique for manipulating the polarization of electromagnetic (EM) waves by shaping the polarization of an antenna into a desired state.  By dynamically adjusting antenna polarization, polarforming enables real-time polarization matching or mismatching with received EM waves, thereby leveraging polarization degrees of freedom (DoFs) to enhance wireless communication performance.  In this article, we first present an overview of the fundamental principles and design approaches underlying the polarforming technique.  We then analyze the key advantages of polarforming, including hardware cost reduction, depolarization mitigation, channel adaptation, signal power enhancement, and interference suppression.  Furthermore, we explore promising applications of polarforming for next-generation wireless networks.  Numerical case studies demonstrate the substantial performance gains of polarforming over conventional fixed-polarization antenna (FPA) systems, along with a discussion of implementation challenges to motivate future research.
\end{abstract}

\section{Introduction}
Polarization characterizes a fundamental property of electromagnetic (EM) waves and offers a new design dimension distinct from conventional time, frequency, and spatial counterparts.  In practical propagation environments, EM waves can be depolarized due to various reasons such as signal reflections, diffractions, scattering, as well as antenna polarization misalignment.  The degree of channel depolarization is typically quantified using cross-polarization discrimination (XPD), defined as the ratio of the average received power in the co-polarized channel to that in the cross-polarized channel~\cite{channel}.  In rich multipath environments, up to six polarization degrees of freedom (DoFs) can theoretically be exploited by deploying six co-located electric and magnetic dipoles with mutually orthogonal polarizations~\cite{survey}.

However, most existing wireless systems with fixed-polarization antennas (FPAs), e.g., linearly polarized antennas (LPAs) and circularly polarized antennas (CPAs), fail to exploit the available polarization DoFs for communication and sensing.  Since channel depolarization is time-varying and environment-dependent, conventional FPA systems cannot fully adapt to channel fluctuations or effectively mitigate depolarization effects.  Among existing solutions, the dual-polarized antenna (DPA)~\cite{dpa} is a commonly adopted approach to leverage polarization DoFs and overcome the limitations of FPAs.  Specifically, each DPA consists of two orthogonally polarized antenna elements, both connected to dedicated radio frequency (RF) chains for independent amplitude and phase control.  Tri-polarized antenna (TPA) systems \cite{tpa} have also gained attention due to the potential to harness additional polarization DoFs in environments with sufficient scattering around the transceiver.  Like DPAs, each TPA incorporates three orthogonally polarized antenna elements, all supported by dedicated RF chains.

While DPAs and TPAs can effectively exploit polarization DoFs, they require significant additional RF hardware resources.  For massive multiple-input multiple-output (MIMO) and extremely large antenna array (ELAA) systems, the benefits of polarization DoFs must be carefully weighed against the considerable hardware cost introduced by these multi-polarized antenna configurations.  Recently, polarforming~\cite{polarforming} has emerged as a novel technique that employs polarization-reconfigurable antennas (PRAs) with phase shifters (PSs)~\cite{polarforming1} to exploit polarization DoFs in wireless communication systems while requiring only a single RF chain per antenna. Moreover, the polarforming antenna~\cite{polarforming2} has been proposed to adaptively control the antenna polarization and tune its position/rotation to effectively explore polarization and spatial DoFs for improved system performance. By integrating both phase shifting and antenna rotation, polarforming allows wireless networks to dynamically reconfigure the polarization of transmit and receive antennas in a cost-effective manner and efficiently leverage polarization DoFs to improve system performance.

Motivated by the significant potential of polarforming for next-generation wireless networks, this article provides a comprehensive overview of its fundamental principles and practical implementation challenges.  To this end, we first present the polarforming framework and its key design approaches.  Next, we analyze the main benefits of polarforming, including reduced hardware costs, robustness to depolarization effects, adaptability to channel variations, enhanced signal power, and improved interference mitigation.  We then explore promising applications of polarforming for future wireless networks.  Finally, we evaluate its performance through numerical case studies and identify the implementation challenges to inspire future research in this emerging field.

\section{Polarforming Framework} \label{sec2}
In this section, we develop a polarforming framework based on wave propagation principles and provide an architecture of polarforming design for the purpose of fully utilizing polarization DoFs.

\begin{figure}[!t]
	\centering
	\fbox{\includegraphics[width=\linewidth]{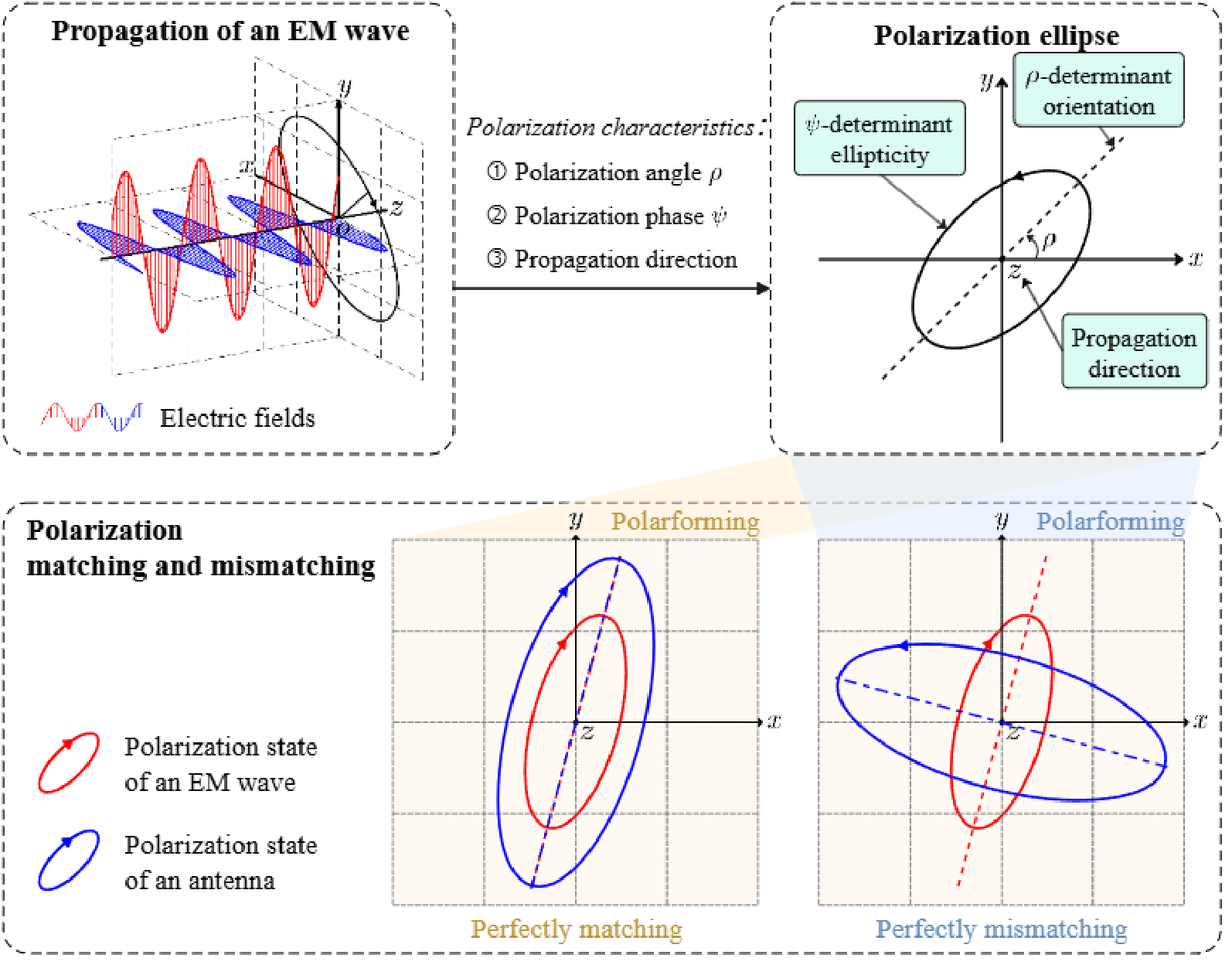}}
	\caption{Illustration of polarforming principles from a propagation perspective.}
	\label{fig_framework}
\end{figure}

\begin{figure}[!t]
	\centering
	\fbox{\includegraphics[width=1\linewidth]{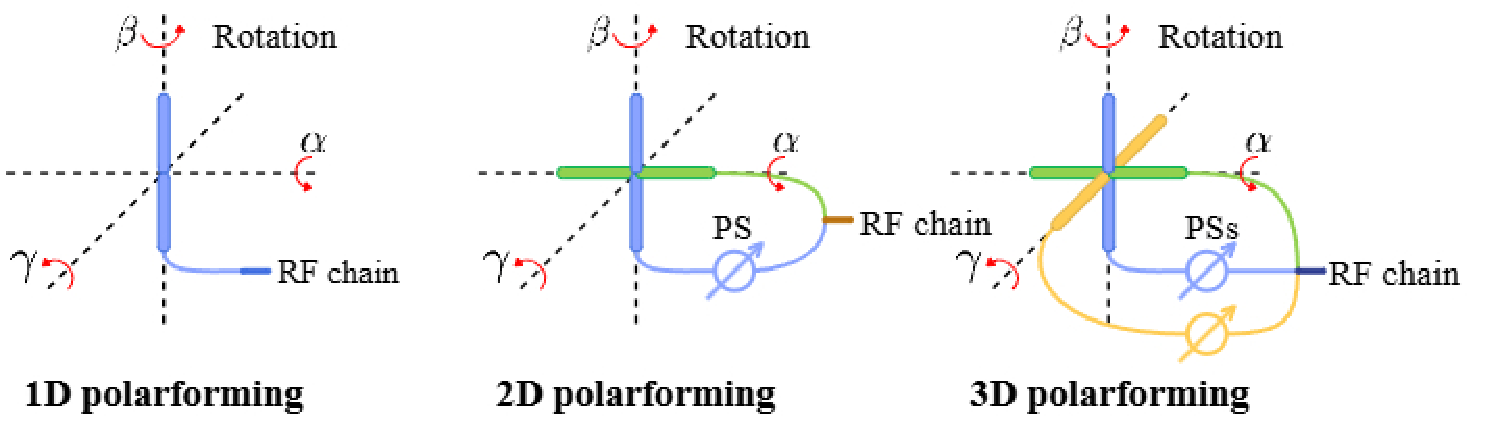}}
	\caption{Antenna architecture of polarforming design by antenna rotation and phase shifting.}
	\label{fig_design}
\end{figure}

\subsection{Physical Principles}
From the perspective of electromagnetism, an EM wave propagating in free space is characterized as a transverse wave.  This means that the radiated electric and magnetic fields have no components along the direction of propagation.  Instead, they are mutually perpendicular and lie on a transverse wavefront orthogonal to the propagation direction.  The magnetic field is coupled to the electric field with a $90^{\circ}$ phase shift, and its magnitude is determined by the wave impedance of the propagation medium.  Once the propagation direction is fixed, only two DoFs remain for the electric field.  These can be effectively represented by a complex vector known as the Jones vector, which characterizes the polarization of the wave.  As illustrated in Fig.~\ref{fig_framework}, the polarization of a wave is governed by the amplitudes and phase difference of its electric field components.  Accordingly, the amplitudes define the polarization angle, while the phase difference determines the polarization ellipticity and handedness.

Polarforming is a specialized form of waveforming that manipulates EM waves in the polarization domain.  It enables wireless systems to dynamically adjust the polarization characteristics of a wave and transform it into a desired polarization state.  As shown in Fig.~\ref{fig_framework}, when the polarization of an antenna matches that of an incoming wave, meaning they have the same orientation, handedness, and ellipticity, the channel power gain reaches its maximum value.  Conversely, exact polarization mismatching occurs when two polarization states are orthogonal, i.e., they have the same ellipticity but orthogonal orientations and opposite handedness, which results in theoretically zero signal reception.

\subsection{Polarforming Design}
The polarforming framework offers a systematic methodology for optimizing wireless systems through flexible polarization adaptation.  Therefore, polarforming can be achieved by modifying the polarization angle via antenna rotation \cite{polarforming2,rotate} and the polarization phase via PSs \cite{polarforming}.  It is worth noting that although this article employs antenna rotation and PSs, polarforming can also be implemented by other possible approaches that allow changing the polarization angle and phase of EM waves.  Fig.~\ref{fig_design} depicts the antenna architecture of polarforming design, which integrates both antenna rotation and phase shifting to enable one-dimensional (1D), two-dimensional (2D), and three-dimensional (3D) polarization control.  Specifically, the design approaches are described below.

\begin{itemize}
	\item[$\bullet$] \textbf{1D polarforming:} This design incorporates only a single antenna element directly connected to an RF chain.  The antenna polarization is adjusted by changing its orientation through mechanical rotation in 3D space, equivalently modifying its polarization angle.  This architecture has also been investigated in the PRA-enabled wireless communication systems \cite{PAA, pama}.
	
	\item[$\bullet$] \textbf{2D polarforming:} This design features two orthogonally polarized antenna elements sharing an RF chain.  A PS in one signal path electronically controls the relative phase between the elements to generate linear, circular, or arbitrary elliptical polarization states~\cite{polarforming}.  To change the polarization angle, the entire antenna can be mechanically rotated in 3D space~\cite{polarforming2}.
	
	\item[$\bullet$] \textbf{3D polarforming:} This design employs three antenna elements with orthogonal polarization orientations, all fed by an RF chain.  By incorporating PSs in two signal paths, the relative phase between any pair of orthogonally polarized signals can be precisely controlled.  Similarly, 3D antenna rotation offers polarization angle adjustment.
\end{itemize}

\section{Polarforming Advantages}
\renewcommand{\arraystretch}{1.3} 
\renewcommand{\multirowsetup}{\centering}
\begin{table*}[!t]
	\centering
	\begin{threeparttable}
		\caption{Comparison of different polarized antenna architectures.}
		\label{tab1}
		\begin{tabular}{|c|c|c|c|c|c|c|}
			\hline
			\multirow{2}{*}{\textbf{Polarization dimension}} 
			& \multirow{2}{*}{\textbf{Polarization scheme}} 
			& \multicolumn{3}{c|}{\textbf{Hardware}} 
			& \multirow{2}{*}{\textbf{Cost}} 
			& \multirow{2}{*}{\textbf{Performance gain}} \\
			\cline{3-5}
			& & \textbf{RF chain} & \textbf{Rotation motor} & \textbf{Phase shifter} & & \\
			\hline
			\multirow{2}{*}{Uni-polarized} & 1D polarforming & Single & \checkmark & \ding{55} & Moderate & Medium \\
			\cline{2-7}
			& LPA & Single & \ding{55} & \ding{55} & Low & Very low \\
			\hline
			\multirow{3}{*}{Dual-polarized} & 2D polarforming & Single & \checkmark & \checkmark (Single) & Moderate & High \\
			\cline{2-7}
			& CPA & Single & \ding{55} & \checkmark (Single) & Low & Low \\
			\cline{2-7}
			& DPA & Double & \ding{55} & \ding{55} & High & High \\
			\hline
			\multirow{2}{*}{Tri-polarized} & 3D polarforming & Single & \checkmark & \checkmark (Double) & Moderate & Very high \\
			\cline{2-7}
			& TPA & Triple & \ding{55} & \ding{55} & High & Very high \\
			\hline
		\end{tabular}
		
		\footnotesize
		LPA: linearly polarized antenna; CPA: circularly polarized antenna; DPA: dual-polarized antenna; TPA: tri-polarized antenna.
	\end{threeparttable}
\end{table*}

We then analyze the key advantages of polarforming in terms of reducing hardware cost, combating channel depolarization, adapting to channel variations, improving signal power, and mitigating interference.

\subsection{Reducing Hardware Cost}
Compared to conventional FPAs, e.g., LPAs and CPAs, and multi-polarized antennas including DPAs and TPAs, Table~\ref{tab1} summarizes the architectural distinctions and competitive performance advantages of polarforming.  In DPA systems, each antenna comprises two orthogonally polarized antenna elements, while TPA systems utilize three orthogonal antenna elements, each element with its dedicated RF chain.  In contrast, the polarforming design requires just one RF chain per antenna, regardless of the number of antenna elements.  This design achieves a twofold or threefold reduction in the number of required RF chains, significantly saving RF hardware costs while maintaining full use of polarization DoFs.  Moreover, employing fewer RF chains reduces overall receiver noise power, which leads to improved system performance, as will be demonstrated by the simulation results in Section~\ref{simu}.

\subsection{Combating Channel Depolarization}
The polarization of EM waves can be notably affected by propagation events, such as reflections, diffractions, and scattering, as well as by system imperfections like antenna polarization misalignment.  In wireless communication systems, the received wave is typically a superposition of line-of-sight (LoS) and non-line-of-sight (NLoS) components.  As a result, the effective polarization of the received wave often deviates from that of the originally transmitted wave, known as channel depolarization.  Polarforming offers a flexible and effective solution by enabling dynamic adjustment of the polarizations of both the transmit and receive PRAs.  Specifically, the transmit PRA can be tuned to ensure the EM wave arrives at the receiver with a desired polarization state, thereby compensating for channel depolarization effects.  Similarly, the receive PRA can dynamically adapt its polarization to match the effective polarization state of the incoming wave.

\subsection{Adapting to Channel Variations}
In practical wireless networks, channel depolarization effects are inherently time-varying, especially for mobile communications.  By dynamically reconfiguring antenna polarization, polarforming allows wireless systems to promptly compensate for channel depolarization, ensuring that the EM wave maintains a desired polarization state.  This adaptive capability is beyond the reach of conventional FPA systems, which lack the flexibility to track variations in the polarized channel.  It is important to note that the effectiveness of polarforming in adapting to channel variations critically depends on the availability of accurate polarized channel state information (CSI), which needs to be estimated efficiently with only one single RF chain per antenna in 1D/2D/3D polarforming.

\subsection{Improving Signal Power}
The impact of channel depolarization is typically overlooked in the design of conventional FPA systems.  That is, while parameters such as beamforming vectors and transmit power are optimized, the polarization remains fixed and unexploited.  This limitation brings significant performance degradation compared to systems with adaptive polarization.  In the worst-case scenario, the polarization of the received wave may be perfectly orthogonal to that of the receive antenna, leading to a complete suppression of received signal power, even under arbitrarily high transmit power.  Polarforming effectively addresses this issue by allowing the transmit and receive antennas to flexibly adjust the antenna polarization.  By aligning the polarization of the received wave with that of the receive antenna, polarforming achieves effective polarization matching, thereby improving the received signal power and reducing the outage probability at the receiver.

\subsection{Mitigating Interference}
In addition to signal power enhancement, polarforming also contributes to interference mitigation. Unlike conventional interference mitigation techniques that operate in the time, frequency, and spatial domains, polarforming behaves in the polarization domain.  It dexterously exploits the polarization differences between the desired signal and the interference.  In this framework, polarforming can align the antenna polarization with the desired signal while keeping orthogonality to the polarization of the interference as much as possible.  This selective polarization matching and mismatching significantly enable wireless systems to separate overlapping signals that may be indistinguishable in other domains.  Furthermore, interference mitigation by polarforming can be jointly applied with the techniques in conventional time, frequency, and spatial domains, thus achieving more efficient interference management in complex wireless environments.

\section{Polarforming Applications}
\begin{figure*}[!t]
	\centering
	\fbox{\includegraphics[width=\linewidth]{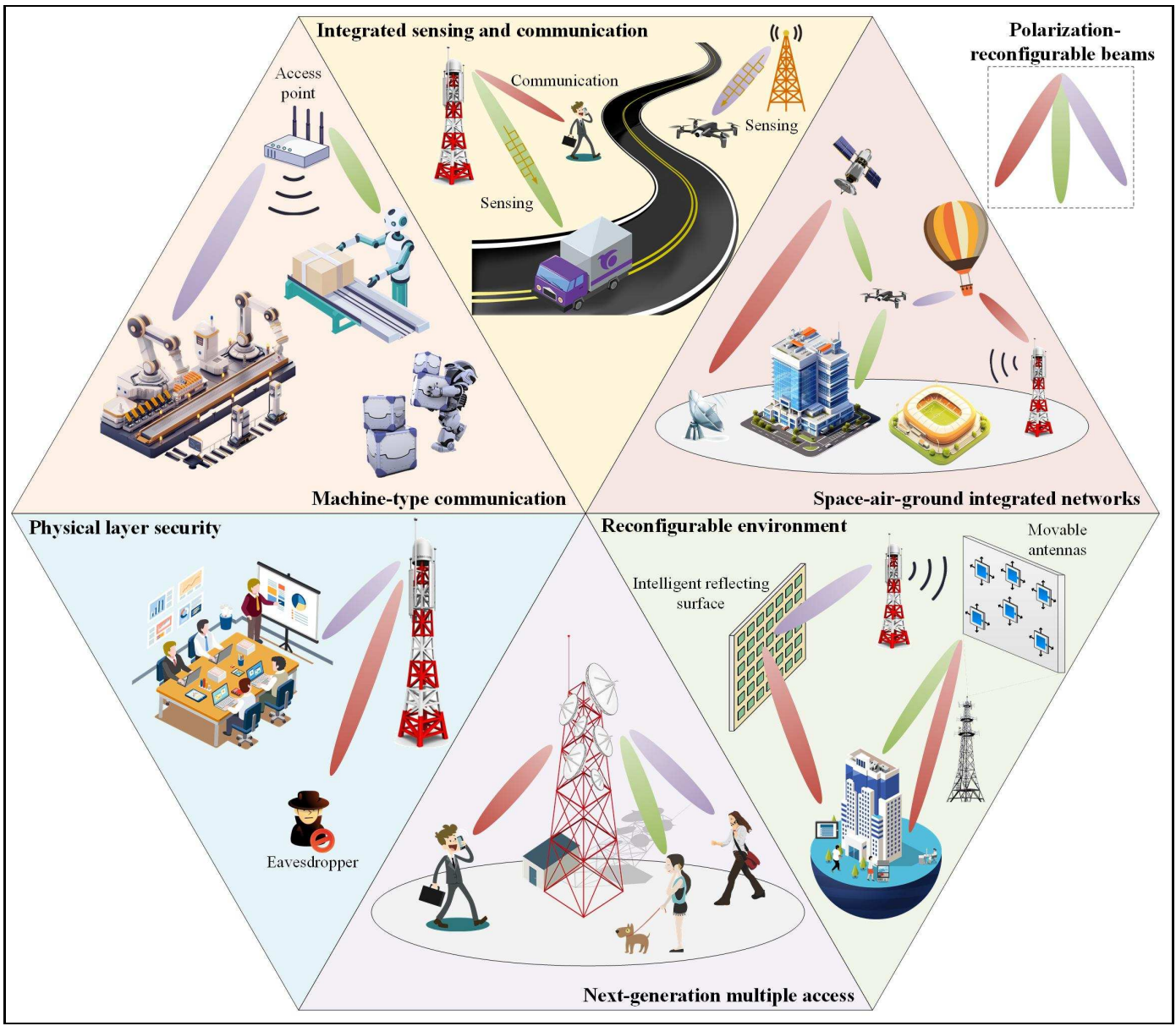}}
	\caption{Typical polarforming applications for wireless networks.}
	\label{applications}
\end{figure*}
Next, we explore possible applications of polarforming for wireless networks in this section, as shown in Fig.~\ref{applications}.

\subsection{Integrated Sensing and Communication}
Integrated sensing and communication (ISAC) is expected to play a critical role in next-generation wireless networks by enabling the joint design and operation of communication and sensing functionalities over shared resources.  For wireless sensing, conventional FPA systems do not account for the polarization change caused by target reflections, leading to a partial loss of information contained in the polarization domain.  In contrast, polarforming, assisted by PRAs, allows for customized antenna polarization, thereby facilitating the capture of complete polarization information from targets and environments.  For ISAC, polarforming enables high-performance and cost-efficient communication and sensing by adapting to the polarized channels of both communication users and sensing targets~\cite{polarforming2}.  In addition, polarforming supports flexible functionality switching between communication and sensing, which can adjust the trade-off between the two tasks by tailoring antenna polarization accordingly.

\subsection{Space-Air-Ground Integrated Networks}
Space-air-ground integrated networks (SAGIN) have emerged as a promising architecture to provide ubiquitous, seamless, and high-capacity wireless connectivity by integrating space-based platforms (e.g., satellites), aerial platforms (e.g., unmanned aerial vehicles (UAVs)), and terrestrial networks.  The appealing gains achieved by polarforming are essential to further enhance the system performance in SAGIN.  For instance, polarization reuse techniques, which are now commonly adopted in satellite broadcasting services~\cite{sata}, simultaneously transmit two independent data streams in two orthogonal polarizations to double the channel capacity.  However, atmospheric propagation effects such as rainfall and ice clouds can alter the polarization of the transmitted EM wave and thus cause it to deviate from the intended polarization by the time it reaches the receive antenna.  The polarization mismatching results in a polarization loss, thereby degrading overall system performance.  By employing polarforming at SAGIN terminals, the antenna polarization can be flexibly adjusted to compensate for the polarization loss inherent in conventional FPA systems, thereby ensuring that the performance gains can be effectively achieved in the polarization domain.  Moreover, the lightweight architecture of polarforming facilitates its deployment on satellites and aerial platforms with limited hardware resources.

\subsection{Reconfigurable Environment}
With the advancement of wireless networks, various reconfigurable technologies, such as intelligent reflecting surface (IRS)~\cite{irs} and movable antenna (MA)/six-dimensional MA (6DMA)~\cite{ma,6dma}, have been rapidly developed.  Polarforming can be integrated into these reconfigurable environments to further enhance wireless connectivity and robustness.  For example, an IRS incorporates a large number of low-cost passive reflecting elements arranged on a planar surface and reconfigures the wireless propagation environment by intelligently adjusting the amplitude and phase of reflected EM waves.  However, reflections inevitably change the polarization of the wave.  By integrating polarforming with IRSs, it becomes possible to jointly control the polarizations of both the incident and reflected waves, thereby unlocking new opportunities for advanced communication scenarios.  Furthermore, MA/6DMA technology leverages spatial DoFs to improve channel power gains through the superposition of multipath components with arbitrary phases.  Similarly, polarforming also considers the signal phase control within the polarization dimension.  The integration of polarforming and MA/6DMA can significantly augment the reconfigurability of wireless environments across both spatial and polarization domains~\cite{polarforming2,6dma}.

\subsection{Next-Generation Multiple Access}
As wireless networks evolve towards 6G and beyond, the demand for massive connectivity and higher spectral efficiency has driven the development of next-generation multiple access (NGMA).  The further exploitation of polarization DoFs, such as polarization multiplexing and polarization diversity, can significantly improve the performance of multiuser systems.  Similar to the application in SAGIN, polarization multiplexing utilizes polarization orthogonality to transmit independent data streams and thus improves the system throughput.  Nevertheless, the propagation characteristics of terrestrial networks differ substantially from those of satellite channels.  In multipath environments, multiple reflections or scattering can severely induce channel depolarization, but it can be mitigated by employing polarforming at the transmitter and/or receiver. On the other hand, with weak channel depolarization, the polarization diversity gain becomes limited due to significant disparities in average signal levels between the co-polarized and cross-polarized branches.  In such cases, polarforming can regulate the polarization of EM waves according to varying depolarization effects, thus maximizing the polarization diversity gain.  Overall, the adaptability enabled by polarforming offers a flexible and effective solution for improving the system performance in NGMA.

\subsection{Physical Layer Security}
Physical layer security (PLS) has long been recognized as a promising paradigm that exploits the inherent properties of wireless channels to deliver security without relying on cryptographic methods.  Polarforming can reduce the risk of eavesdropping on legitimate information by exploiting polarization orthogonality.  In particular, for legitimate users, polarforming can match the receive antenna to the polarization of the incoming wave, thereby maximizing the receive signal-to-interference-plus-noise ratio (SINR).  For eavesdroppers, polarforming can intentionally misalign the polarization of the EM wave with that of the eavesdroppers' antennas.  In theory, cooperative polarforming design among legitimate terminals can achieve perfect polarization mismatching between the EM wave reaching the eavesdroppers and their receive antennas.  In other words, no signal can be received by eavesdroppers, i.e., absolute security.  Besides, when the legitimate user and the eavesdropper are in close proximity, information leakage remains inevitable, even with secure beamforming in conventional FPA systems.  In such cases, polarforming can still effectively distinguish between the legitimate user and the eavesdropper based on polarization differences.  Therefore, polarization is a highly desirable and valuable resource dimension that can be effectively explored through polarforming to enhance PLS in future wireless networks.

\subsection{Machine-Type Communication}
With the ongoing development of the Internet of Things (IoT), machine-type communication (MTC) has become a key component in enabling smart cities, intelligent industries, and smart homes.  Due to the dense deployment of IoT devices in confined areas, the signal propagation often undergoes multiple reflections and scattering.  As a result, the polarization of EM waves is often altered during transmission.  This polarization alteration can lead to signal degradation, increased interference, and reduced system reliability.  Polarforming offers a promising solution to these challenges by dynamically reconfiguring antenna polarization in response to varying channel polarization conditions.  By adapting the polarization of the antenna to match that of the desired signal, polarforming improves received signal power.  Simultaneously, it can suppress interference from undesired sources by introducing polarization mismatching.  This dual capability effectively improves link reliability and spectral efficiency in dense MTC environments.

\section{Performance Evaluation and Challenges}
In this section, we present simulation results to evaluate the performance of polarforming and compare it with the conventional benchmark schemes.  In addition, we discuss the open challenges that need to be considered in order to unlock the full potential of polarforming for future wireless networks.

\subsection{Simulation Results}\label{simu}
We investigate the achievable rate of a MIMO system equipped with $N$ transmit and $M$ receive PRAs using the polarforming design in Fig.~\ref{fig_design}.  Unless otherwise specified, we set $M=N=6$.  In the simulations, we adopt the polarized channel model in~\cite{HWS22} and obtain the optimal phase shifts/antenna rotation for polarforming by exhaustive search for simplicity~\cite{polarforming}.  The SNR is defined as the ratio of the average received signal power to noise power.

\begin{figure}[!t]
	\centering
	\fbox{%
		\begin{tabular}{c}
			\subfloat[][]{\label{1D}\includegraphics[width=\columnwidth]{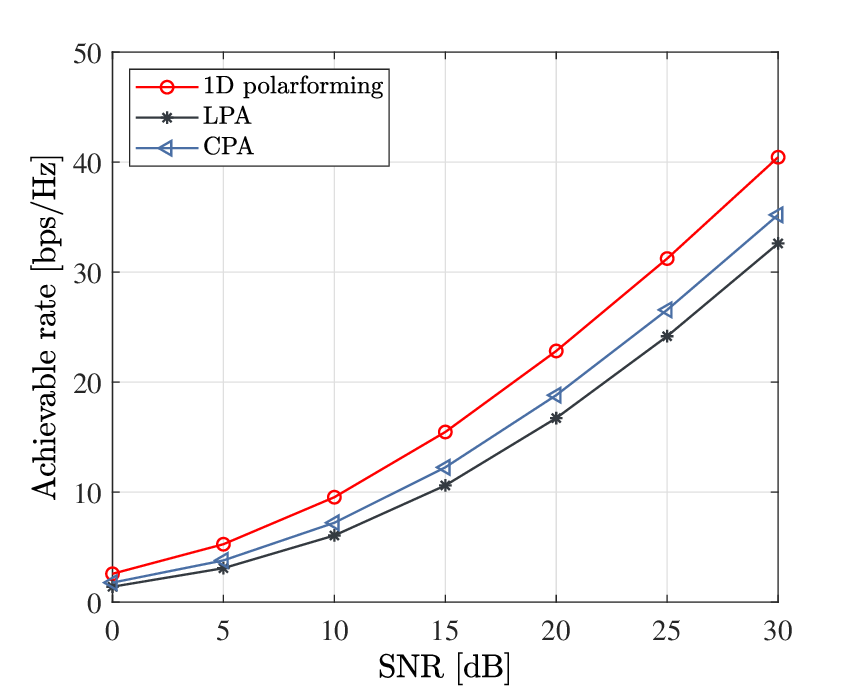}} \\
			\subfloat[][]{\label{2D}\includegraphics[width=\columnwidth]{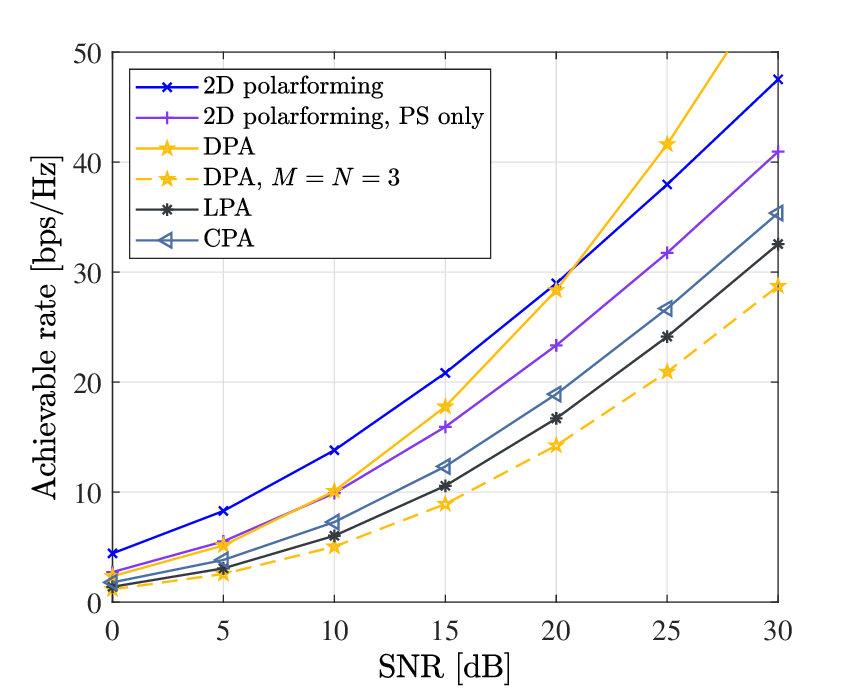}} \\
			\subfloat[][]{\label{3D}\includegraphics[width=\columnwidth]{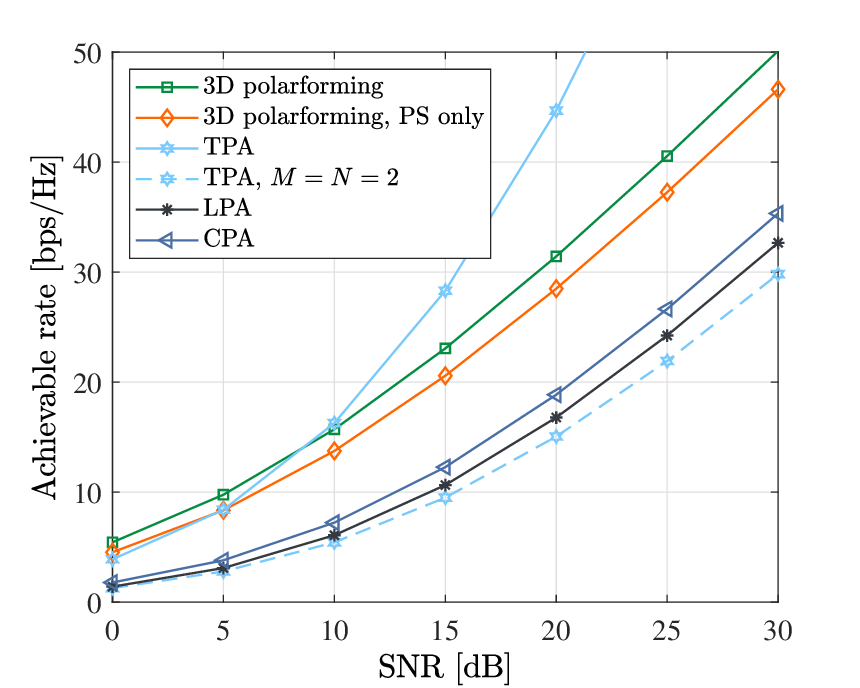}}
		\end{tabular}%
	}
	\caption{Achievable rate of polarforming over conventional schemes.}
	\label{SNR}
\end{figure}

Fig.~\ref{SNR} illustrates the achievable rate versus SNR for different schemes with the inverse XPD being 0 dB.  In Fig.~\ref{SNR}\subref{1D}, 1D polarforming consistently outperforms the LPA and CPA schemes due to its flexible polarization adjustment capability.  Besides, as shown in Fig.~\ref{SNR}\subref{2D}, 2D polarforming achieves notably better performance than the DPA scheme with $M=N=3$.  This is because the DPA scheme employs only half the number of antennas compared to the polarforming scheme under the same number of RF chains.  Similar results can also be observed in comparison to the TPA scheme with $M=N=2$ in Fig.~\ref{SNR}\subref{3D}, where each antenna requires three dedicated RF chains.  Moreover, the polarforming scheme in Figs.~\ref{SNR}\subref{2D} and \ref{SNR}\subref{3D} using only PSs without antenna rotation brings a tolerable performance loss, but significantly reduces the hardware cost by eliminating the mechanical rotation module, thus achieving a trade-off between performance gain and implementation complexity.  Furthermore, it can be observed in Figs.~\ref{SNR}\subref{2D} and \ref{SNR}\subref{3D} that the DPA and TPA schemes with the same number of antennas as the polarforming scheme attain higher achievable rates at high/moderate SNR; however, their performance falls below that of polarforming in the moderate/low SNR regime.  This is because, under the same number of antennas, the DPA and TPA schemes have higher multiplexing gains enabled by additional RF chains.  Nevertheless, in low-SNR scenarios, the larger number of RF chains causes increased noise power at the receiver, thereby offsetting the performance gains. 

\begin{figure}[!t]
	\centering
	\fbox{\includegraphics[width=1\linewidth]{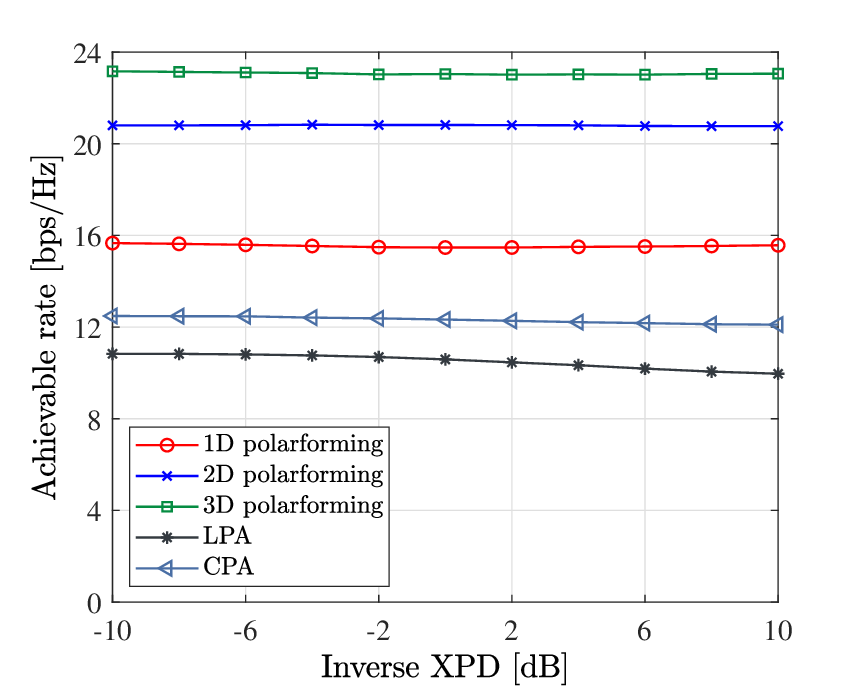}}
	\caption{Impact of channel depolarization on the performance of polarforming.}
	\label{xpd}
\end{figure}

Fig.~\ref{xpd} investigates the impact of channel depolarization on the performance of polarforming, where the SNR is set to 15 dB.  In general, a system that maintains a stable rate across a wide range of inverse XPD values is considered more robust to channel depolarization.  It can be seen that the achievable rate of the polarforming scheme remains nearly constant with varying inverse XPD, while the CPA and LPA schemes experience performance degradation as the inverse XPD increases. This is because both the CPA and LPA schemes are based on FPAs, and thus cannot dynamically adapt to varying levels of channel depolarization.  In addition, the performance of 3D polarforming surpasses that of 2D polarforming, which in turn outperforms 1D polarforming.  This indicates that the polarization adjustment in higher dimensions can more effectively mitigate depolarization effects, at the cost of increased hardware complexity.

\subsection{Challenges}
\subsubsection{Polarized Channel Estimation}\label{csi}
Accurate estimation of the CSI in the polarization domain is a fundamental prerequisite for fully unleashing the performance gains offered by polarforming. Unlike conventional DPA and TPA systems, where each antenna element is assigned a dedicated RF chain, polarforming employs only a single RF chain to serve all elements within one antenna. Consequently, conventional channel estimation techniques may become ineffective under such hardware constraints. In addition, channel variations induced by the movement of terminals or scatterers make the measurement and modeling of polarized channels more challenging. The development of efficient polarized channel estimation algorithms is thus a promising direction for future research. 

\subsubsection{Implementation Issues}
The practical implementation of polarforming poses several challenges at the hardware level, particularly in antenna design and rotation control.  Ideally, a polarized antenna should confine the oscillating electric field strictly to its designated orientation, with no energy leaking into the orthogonal orientation.  However, in practical antenna design and fabrication, the imperfections in the antenna architecture inevitably lead to cross-polarization leakage, which is commonly characterized by cross-polarization isolation (XPI).  This leakage intensifies channel depolarization, which in turn degrades the performance of polarforming in exploiting polarization gains. Furthermore, polarforming via antenna rotation requires the integration of extra rotation modules into existing systems. The joint control and scheduling of both the communication (e.g., phase shifts) and rotation modules require further investigation in future work.

\section{Conclusion}
In this article, the promising concept of polarforming was introduced, which highlights its great potential for enhancing wireless network performance.  We thoroughly discussed the physical principles, design approaches, performance advantages, and promising applications of polarforming.  By flexibly adjusting antenna polarization, polarforming can achieve either polarization matching or mismatching, thereby allowing more effective utilization of polarization DoFs compared to conventional DPA/TPA and FPA systems.  As polarforming remains in its early stage of research, there are many open challenges and opportunities to unlock its full potential.   Therefore, more research efforts are necessary to tackle these challenges and make polarforming a practical and effective solution for future wireless networks.

\vspace{0.2cm}
Jingze Ding (djz@stu.pku.edu.cn) is a Ph.D. candidate with the School of Electronics, Peking University.

\vspace{0.2cm}
Zijian Zhou (zijianzhou@link.cuhk.edu.cn) is a Ph.D. candidate with the School of Science and Engineering, The Chinese University of Hong Kong, Shenzhen.

\vspace{0.2cm}
Xiaodan Shao (x6shao@uwaterloo.ca) is a Postdoctoral Fellow with the Department of Electrical and Computer Engineering, University of Waterloo.

\vspace{0.2cm}
Bingli Jiao (jiaobl@pku.edu.cn) is a professor with the School of Computing and Artificial Intelligence, Fuyao University of Science and Technology. He is also with the School of Electronics, Peking University.

\vspace{0.2cm}
Rui Zhang [F'17] (elezhang@nus.edu.sg) is a professor with the School of Science and Engineering, The Chinese University of Hong Kong, Shenzhen, and Shenzhen Research Institute of Big Data. He is also with the ECE Department of National University of Singapore.
\end{document}